\documentstyle{article}
\begin{document}
\title{The coplanar $n$-body gravitational lens :\\ a general theorem}
\author{T. Richard Carson\\
School of Physics and Astronomy, University of St Andrews\\
North Haugh, St Andrews, Scotland, UK, KY169SS\\
email: trc@st-andrews.ac.uk}
\maketitle
\begin{abstract}

Generalizing a result for the binary lens, similar 
alternative expressions are also given for the Jacobian determinant for
a gravitational lens consisting of an arbitrary number of  
discrete lensing centres, with arbitrary masses and locations in a plane.
The results for a binary lens are recovered, agreeing with some already
known results, and correcting errors in others.

\end{abstract}

\section{Introduction}

Since the earliest studies of Einstein (1915,1916) on the deflection of
light by gravitational fields in general relativity, a vast research and
associated publication record in gravitational lensing, 
 both observational and theoretical, has blossomed. The simple single
mass point lens model is already the main paradigm for the understanding of 
the many ($\sim 10^{2}$) lensing events detected by the various monitoring
and follow-up programmes under way. Binary lenses are also the focus of
much effort, not only for their own intrinsic interest, but also for their
potential as a tool, particularly in the detection of planetary 
systems. More sophisticated lens models, including many-body systems, are of 
interest too because of their possible galactic and cosmological applications.

\section{The lens equation(s)}

In an orthogonal rectangular co-ordinate system (x,y,z), with the origin 
at the observer, and the z-axis as optical axis, let a source be
located in a (source) plane S at (angular size) distance $D_{\rm S}$, and 
let there be $n$ masses located in a (lens) plane L at distance $D_{\rm L}$.
Rays from a source point at angular co-ordinates ($x_{s},y_{s}$),
on crossing the lens plane are so deflected as to produce an image
point (of which there may be more than one) at angular co-ordinates
($x_{i},y_{i}$). The lensing conditions for small angles, see for example
Paczy\'{n}ski (1996), are given by:
\[x_{i} - x_{s} = \alpha _{x} (D_{\rm S} - D_{\rm L})/D_{\rm S}; \,\,\,
  y_{i} - y_{s} = \alpha _{y} (D_{\rm S} - D_{\rm L})/D_{\rm S} \]
where $\alpha _{x}$ and $\alpha _{y}$ are the angular deflections in the
directions of the x and y axes respectively. In a two-dimensional 
vector notation, with $\vec {r} = (x,y)$ and $\vec {\alpha } = 
(\alpha _{x},\alpha _{y})$ the above conditions may be written 
as the single equation :
\[\vec {r}_{i} - \vec {r}_{s} = [(D_{\rm S} - D_{\rm L})/D_{\rm S}]
\vec {\alpha } \]
For a spherically symmetric mass $M$ the Einstein (1916) gravitational 
deflection is $\alpha = 4GM/c^{2}R_{0}$, where $R_{0}$ is the radial distance
of closest approach or impact parameter. In the (rectilinear) ray optics
approximation, the image position is deemed to be where the ray crosses
the lens plane. Thus for each lensing mass $M_{l}$ located at $\vec {r}_{l}$
($l=1,n$), setting $\vec {r}_{il} = |\vec {r}_{i} - \vec {r}_{l}|$,
$r_{il} = |\vec {r}_{il}|$, and $R_{0} = D_{L}r_{il}$,
the contribution to the deflection is $\vec {\alpha }_{l} = 
\alpha _{0} (\vec {r}_{i} - \vec {r}_{l})/r_{il}^{2}$ where
$\alpha _{0} = 4GM_{l}/c^{2}D_{\rm L}$.
The lens equation for the $n$-body system then takes the form:
\[\vec {r}_{i} - \vec {r}_{s} = [(D_{\rm S} - D_{\rm L})/D_{\rm S}]
\sum _{l=1}^{n}\vec {\alpha }_{l} =
\sum _{l=1}^{n} m_{l}(\vec {r}_{i} - \vec {r}_{l})/r_{il}^{2} \]
with $m_{l} = [(D_{\rm S} - D_{\rm L})/D_{\rm S}D_{\rm L}]4GM_{l}/c^{2} = 
(M_{l}/M)r_{\rm E}^{2}$, where $M = \sum M_{l}$ while 
$r_{\rm E} = [4GMc^{-2}(D_{\rm S}-D_{\rm L})/D_{\rm S}D_{\rm L}]^{1/2}$ is the
{\em Einstein Ring} angular radius for mass $M$. Without any loss of generality
one may set $m = \sum m_{l} = 1$, thus expressing all the angular measures in 
units of $r_{\rm E}$. The lens equation may alternatively be written (Witt,1990) 
in complex variable notation with $z = x + iy$ and $\bar {z} = x - iy$ as :
\[z_{i} - z_{s} = \sum _{l=1}^{n} m_{l}(z_{i} - z_{l})/|z_{i} -z_{l}|^{2} = 
\sum _{l=1}^{n}m_{l}/(\bar {z}_{i} - \bar {z}_{l}) \]

\section{Image amplification}

Since the intensity of radiation is invariant along a ray the brightness 
of the lensed source relative to that of the unlensed source (the 
magnification or amplification) is the ratio of the areas or solid
angles subtended at the observer, given by $A = J^{-1}$, where $J$
is the Jacobian (determinant) of the transformation mapping $\vec {r}_{i}$
into $\vec {r}_{s}$ :
\[J = |\partial \vec {r}_{s}/\partial \vec{r}_{i}| = 
      |\partial (x_{s},y_{s})/\partial (x_{i},y_{i})| = 
      |\delta _{xy} - a_{xy}| \]
with $a_{xx} = -a_{yy}$, $a_{xy} = a_{yx}$ and :
\[a_{xx} = \sum _{l}m_{l}\rho _{l}^{-4}(\eta_{l}^{2}-\xi _{l}^{2});\,\,\,
a_{xy} = 2 \sum _{l}m_{l}\rho _{l}^{-4}\xi _{l}\eta_{l} \]
where $\xi _{l} = x_{i}-x_{l}$, $\eta_{l} = y_{i}-y_{l}$,
and $\rho _{l} = r_{il}$. Then :
\[J = (1 - a_{xx})(1 + a_{xx}) - a_{xy}^{2} =
1 - a_{xx}^{2} - a_{xy}^{2} = 1 - K \]
\[K = [\sum _{l}m_{l}\rho _{l}^{-4}(\eta_{l}^{2}-\xi _{l}^{2})]^{2} + 
[2\sum _{l}m_{l}\rho _{l}^{-4}\xi _{l}\eta _{l}]^{2} \]
Writing $K = K_{1}+K_{2}$, expanding each term, and separating the 'diagonal' 
and 'non-diagonal' parts of each, gives $K_{1}=K_{11}+K_{12}$ and
$K_{2}=K_{21}+K_{22}$ where :
\[K_{11}=\sum _{l}[m_{l}\rho _{l}^{-4}(\eta_{l}^{2}-\xi _{l}^{2})]^{2};\,\,\,
  K_{21}=\sum _{l}[2m_{l}\rho _{l}^{-4}\xi _{l}\eta _{l}]^{2} \]
\[K_{12}=2\sum _{j\neq k}\sum _{k}m_{j}m_{k}\rho _{j}^{-4}\rho _{k}^{-4}
(\eta _{j}^{2}-\xi _{j}^{2})(\eta _{k}^{2}-\xi _{k}^{2}) \]
\[K_{22}= 8\sum _{j\neq k}\sum _{k}m_{j}m_{k}\rho _{j}^{-4}\rho _{k}^{-4}
\xi _{j}\eta _{j}\xi _{k}\eta _{k} \]
Adding the 'diagonal' terms $K_{11}$ and $K_{21}$ gives :
\[K_{3}=[\sum _{l}m_{l}\rho _{l}^{-2}]^{2}-
2\sum _{j\neq k}\sum _{k}m_{j}m_{k}\rho _{j}^{-2}\rho _{k}^{-2} =K_{31}+K_{32} \]
where $K_{32}$ may also be written :
\[K_{32}=-2\sum _{j\neq k}\sum _{k}m_{j}m_{k}\rho _{j}^{-4}\rho _{k}^{-4} 
  (\xi _{j}^{2}+\eta _{j}^{2})(\xi _{k}^{2}+\eta _{k}^{2}) \]
It then follows that :
\[K_{12}+K_{32}=-4\sum _{j\neq k}\sum _{k}m_{j}m_{k}\rho _{j}^{-4}\rho _{k}^{-4} 
  [\xi _{j}^{2}\eta _{k}^{2}+\xi _{k}^{2}\eta _{j}^{2}] \]
whereupon, adding the remaining 'non-diagonal' term $K_{22}$ gives
$K_{4}=K_{12}+K_{22}+K_{32}$ with :
\[K_{4}=-4\sum _{j\neq k}\sum _{k}m_{j}m_{k}\rho _{j}^{-4}\rho _{k}^{-4}
  [\xi _{j}\eta _{k}-\xi _{k}\eta _{j}]^{2} \]
so that finally $K = K_{31}+K_{4}$ or :
\[K=[\sum _{l}m_{l}\rho _{l}^{-2}]^{2}-4\sum _{j\neq k}\sum _{k}m_{j}m_{k}\rho _{j}^{-4}\rho _{k}^{-4}
  [\xi _{j}\eta _{k}-\xi _{k}\eta _{j}]^{2} \]
The same result may naturally be obtained, perhaps more easily,
from the complex variable formulation of the lens equation.
There it may be shown (Witt,1990) that :
\[J = (\partial z_{s}/\partial z_{i})^{2} -
      |\partial z_{s}/\partial \bar{z}_{i}|^{2} =
  1 - |\sum _{l}m_{l}\bar {\zeta }_{l}^{-2}|^{2} \]
where $\zeta _{l}=z_{i}-z_{l}$ and $\rho _{l}=|\zeta _{l}|$.
Thus, with $K=J-1$ :
\[K=[\sum _{j}m_{j}\zeta _{j}^{-2}][\sum _{k}m_{k}\bar {\zeta }_{k}^{-2}]
=\sum _{j}\sum _{k}m_{j}m_{k}[\zeta _{j}\bar {\zeta }_{k}]^{-2} \]
Expanding and separating the 'diagonal' and 'non-diagonal' parts gives
$K=K_{5}+K_{6}$ with :
\[K_{5} = \sum _{l}m_{l}^{2}\rho _{l}^{-4} = 
[\sum _{l}m_{l}\rho _{l}^{-2}]^{2} - 2 \sum _{j\neq k}\sum _{k}
m_{j}m_{k}\rho _{j}^{-2}\rho _{k}^{-2} \]
\[K_{6}=\sum _{j\neq k}\sum _{k}m_{j}m_{k}
[\zeta _{j}^{-2}\bar {\zeta }_{k}^{-2}+\zeta _{k}^{-2}\bar {\zeta }_{j}^{-2}] \]
Putting $K_{5}=K_{51}+K_{52}$, where $K_{51}=K_{31}$, the second term $K_{52}$ may be rewritten :
\[K_{52} = -2\sum _{j\neq k}\sum _{k}m_{j}m_{k}\rho _{j}^{-4}\rho _{k}^{-4}
[\zeta _{j}\bar {\zeta }_{j}\zeta _{k}\bar {\zeta }_{k}] =K_{32} \]
while $K_{6}$ may be rewritten as :
\[K_{6}=\sum _{j\neq k}\sum _{k}m_{j}m_{k}\rho _{j}^{-4}\rho _{k}^{-4}
[\bar {\zeta }_{j}^{2}\zeta _{k}^{2}+\zeta _{j}^{2}\bar {\zeta }_{k}^{2}] \]
Now adding the 'non-diagonal' terms $K_{52}$ and $K_{6}$ gives :
\[K_{7}=\sum _{j\neq k}\sum _{k}m_{j}m_{k}\rho _{j}^{-4}\rho _{k}^{-4}
[\bar {\zeta }_{j}\zeta _{k}-\zeta _{j}\bar {\zeta }_{k}]^{2} = K_{4} \]
since $[\bar {\zeta }_{j}\zeta _{k}-\zeta _{j}\bar {\zeta }_{k}] = 
  2 i [\eta _{j}\xi _{k}-\xi _{j}\eta _{k}]$
thus giving the result $K = K_{51}+K_{7} = K_{31}+K_{4}$ as before.
It may be remarked here that 
$[\xi _{j}\eta _{k}-\xi _{k}\eta _{j}] = [\vec {r}_{ij} \times \vec {r}_{ik}]$
so that :
\[K_{4} = -4\sum _{j\neq k}\sum _{k}m_{j}m_{k}
[\nabla _{i}(r_{ij}^{-1}) \times \nabla _{i}(r_{ik}^{-1})]^{2}\]
where the two-dimensional gradient $\nabla _{i} = 
\partial /\partial \vec {r}_{i}$.
 \section{The general binary lens}

In the case $n = 2$ the double summation reduces to a single term
with $j=1,k=2$ giving :
\[J=1-[m_{1}\rho _{1}^{-2}+m_{2}\rho _{2}^{-2}]^{2}
+4m_{1}m_{2}\rho _{1}^{-4}\rho _{2}^{-4}[C_{12}]^{2} \]
where $C_{12}=[\xi _{1}\eta _{2}-\xi _{2}\eta _{1}]$ now reduces to :
\[C_{12}=[(x_{1}y_{2}-x_{2}y_{1})+x_{i}(y_{1}-y_{2})-y_{i}(x_{1}-x_{2})] \]
which may be simplified by an appropriate choice of co-ordinates.
Taking the origin anywhere on the line joining the two masses,
then $y_{1}/x_{1}=y_{2}/x_{2}$ and $x_{1}y_{2}-x_{2}y_{1}=0$ gives
$C_{12}=[(x_{2}-x_{1})y_{i}-(y_{2}-y_{1})x_{i}]$ which is the same function of
$x_{i}$ and $y_{i}$ for given values of $x_{2}-x_{1}=2x_{0}$ and
$y_{2}-y_{1}=2y_{0}$, namely $C_{12}=2[x_{0}y_{i}-y_{0}x_{i}]$. The further simplification of setting $y_{0}=0$, so that the x-axis lies along the
line joining the two masses, with $y_{1}=y_{2}=0$, gives $C_{12}=2x_{0}y_{i}$.
There still remains the choice of the value of $x_{1}$ or $x_{2}$, for which 
there are three 'natural' possibilities : Case (a) - Origin mid-way between the
masses with $x_{1}=-x_{0},x_{2}=+x_{0}$; Case (b) - Origin at mass $M_{1}$ with
$x_{1}=0, x_{2}=+2x_{0}$; Case (c) - Origin at mass $M_{2}$ with $x_{1}=-2x_{0},
x_{2}=0$.

\subsection{The critical curves}

The critical curves, or loci of infinite amplification, are given by
the condition $J=0$, which on rationalising takes the form $D_{12}=0$ with :
\[D_{12}=\rho _{1}^{4}\rho _{2}^{4}J
=\rho _{1}^{4}\rho _{2}^{4}-[m_{1}\rho _{2}^{2}+m_{2}\rho _{1}^{2}]^{2}
+16m_{1}m_{2}x_{0}^{2}y_{i}^{2} \]
in agreement with Schneider \& Weiss (1986). Adopting plane polar co-ordinates
 $\vec {r}_{i} = (r_{i}, \theta _{i})$, then $D_{12} = \sum _{m} a_{m}
\cos ^{m}(\theta _{i})$ where the coefficients $a_{m}$ depend on the choice
of origin :

Case (a) - $\,\,\,\vec{r}_{i}=(r_{0},\theta _{0})$ :
\[a_{0}=(r_{0}^{2}+x_{0}^{2})^{4}-(r_{0}^{2}+x_{0}^{2})^{2}(m_{1}+m_{2})^{2}
+16m_{1}m_{2}r_{0}^{2}x_{0}^{2} \]
\[a_{1}=4r_{0}x_{0}(r_{0}^{2}+x_{0}^{2})(m_{1}^{2}-m_{2}^{2}) \]
\[a_{2}=-4r_{0}^{2}x_{0}^{2}[2(r_{0}^{2}+x_{0}^{2})^{2}+(m_{1}+m_{2})^{2}] \]
\[a_{3}=0;\,\,\,a_{4}=16r_{0}^{4}x_{0}^{4} \]
In the symmetric case where $m_{1}=m_{2}=1/2$, $a_{1}=0$, the term in
$\cos \theta _{i}$ drops out, and the equation reduces to a quadratic equation
in $\cos ^{2}\theta _{i}$, as given by Equation 9a of Schneider \& Weiss (1986),
with the solution :
\[\cos ^{2}\theta _{0}=[1+2(r_{0}^{2}+x_{0}^{2})^{2}+
\{1+8(r_{0}^{4}+x_{0}^{4})\}^{1/2}]/(8r_{0}^{2}x_{0}^{2}) \]

Case (b) - $\,\,\,\vec {r}_{i}=(r_{1},\theta _{1})$, with $r_{1}=\rho _{1}$ :
\[a_{0}=(r_{1}^{2}+4x_{0}^{2})^{2}(r_{1}^{4}-m_{1}^{2})
-2r_{1}^{2}(r_{1}^{2}+4x_{0}^{2})m_{1}m_{2} \]
\[\,\,\,+16r_{1}^{2}x_{0}^{2}m_{1}m_{2}-r_{1}^{4}m_{2}^{2} \]
\[a_{1}=8r_{1}x_{0}[r_{1}^{2}m_{1}m_{2}
-(r_{1}^{2}+4x_{0}^{2})(r_{1}^{4}-m_{1}^{2})] \]
\[a_{2}=16r_{1}^{2}x_{0}^{2}[r_{1}^{4}-m_{1}^{2}-m_{1}m_{2}] \]

Case (c) - $\,\,\,\vec {r}_{i}=(r_{2},\theta _{2})$, with $r_{2}=\rho _{2}$ :
\[a_{0}=(r_{2}^{2}+4x_{0}^{2})^{2}(r_{2}^{4}-m_{2}^{2})
-2r_{2}^{2}(r_{2}^{2}+4x_{0}^{2})m_{1}m_{2} \]
\[\,\,\,+16r_{2}^{2}x_{0}^{2}m_{1}m_{2}-r_{2}^{4}m_{1}^{2} \]
\[a_{1}=8r_{2}x_{0}[(r_{2}^{2}+4x_{0}^{2})(r_{2}^{4}-m_{2}^{2})-r_{2}^{2}m_{1}
m_{2}] \]
\[a_{2}=16r_{2}^{2}x_{0}^{2}[r_{2}^{4}-m_{2}^{2}-m_{1}m_{2}] \]
from which it will be evident that Case (b) and Case (c) are equivalent
with respect to the interchange of the labels 1 and 2 together with
the change of sign of $x_{0}$. Both therefore correct several errors in
Equation 9b of Schneider and Weiss (1986), after allowing for their
relative disposition of the binary masses.
In the limit $x_{0}=0$ the equations for all three cases
degenerate to $r_{i}^{4}-(m_{1}+m_{2})^{2}=0$, giving
$r_{i}=(m_{1}+m_{2})^{1/2}$, so that the critical curve
is the Einstein Ring for the combined masses of the binary.
On the other hand in the limit $x_{0}=\infty$, where either $\rho _{1}=\infty$
or $\rho _{2}=\infty$, the condition $J=0$ reduces to
$\rho _{l}^{4}-m_{l}^{2}=0, (l=1,2)$, with the solution $\rho _{l}=m_{l}^{1/2}$
representing the now detached Einstein Rings around each mass.

\end{document}